\documentclass[prb,aps,onecolumn,showpacs,floats,a4paper,times,12pt]{revtex4}
\usepackage{epsfig,bm,epsf,graphics}
\begin{document}
\title{First-Order Transition in XY Fully Frustrated Simple Cubic Lattice}
\author{V. Thanh Ngo$^{a,b}$, D. Tien Hoang$^{a}$ and H. T. Diep\footnote{ Corresponding author, E-mail:diep@u-cergy.fr }}
\address{Laboratoire de Physique Th\'eorique et Mod\'elisation,
Universit\'e de Cergy-Pontoise, CNRS UMR 8089\\
2, Avenue Adolphe Chauvin, 95302 Cergy-Pontoise Cedex, France\\
$^a$ Institute of Physics, P.O. Box 429,   Bo Ho, Hanoi 10000,
Vietnam\\
$^b$ Division of Fusion and Convergence of Mathematical
Sciences, National Institute for Mathematical Sciences, Daejeon, Republic of
Korea}

\begin{abstract}
We study the nature of the phase transition in the fully frustrated simple cubic lattice with the XY spin model.  This system is the Villain's model generalized in three dimensions.  The ground state is very particular with a 12-fold degeneracy.  Previous studies have shown unusual critical properties.  With the powerful
Wang-Landau flat-histogram Monte Carlo method, we carry out in this work intensive simulations with very large lattice sizes.  We show that the phase transition is clearly of first order, putting an end to the uncertainty which has lasted for more than twenty years.
\end{abstract}

\maketitle

\section{Introduction}

One of the most fascinating tasks of statistical physics is the study of the phase transition in systems of interacting particles.  Much progress has been made in this field since 50 years.  Finite-size theory, renormalization group analysis, numerical simulations, ... have contributed to the advance of the knowledge on the phase transition.  Exact methods have been devised to solve with mathematical elegance many problems in two dimensions. But as improvements are progressing, new and more complicated challenges also come in from new discoveries of materials and new applications.  Renormalization group which has predicted with success critical behaviors of ferromagnets has many difficulties to deal with frustrated systems. Numerical simulations which did not need huge memory and long calculations for simple systems require now new devices, new algorithms to improve convergence in systems with extremely long relaxation time, or in systems whose microscopic states are difficult to access.  One class of these systems is called 'frustrated systems' introduced in the seventies in the context of spin glasses.   These frustrated systems are very unstable due to the competition between different kinds of interaction. However they are periodically defined (no disorder) and therefore subject to exact treatments.  This is the case of several models in two dimensions\cite{Diep-Giacomini}, but in three dimensions frustrated systems are far from being understood even on basic properties such as the order of the phase transition (first or second order, critical exponents, ...).   Let us recall the definition of a frustrated system.
When a spin cannot fully satisfy energetically all the interactions with its neighbors, it is
"frustrated".   This occurs when the interactions are in competition with
each other or when the lattice geometry does not allow to satisfy all interaction bonds simultaneously. A well-known example is the stacked triangular antiferromagnet with interaction between nearest-neighbors.

The frustration
in spin systems causes many unusual properties such as large ground state (GS)
degeneracy, successive phase transitions
with complicated nature, partially disordered phase, reentrance and disorder lines.
Frustrated systems still constitute at present a challenge for investigation methods.  For  recent reviews, the
reader is referred to Ref. \onlinecite{Diep2005}.

In this work, we are interested in the nature of the phase transition of the classical XY spin model
in the fully frustrated simple cubic lattice (FFSCL) shown  in Fig. \ref{fig:SCFF}.   Although phase transition in strongly frustrated systems has been a
subject of intensive investigations in the last 20 years, many aspects are still not understood at present. One of the most studied systems is the stacked triangular antiferromagnet (STA) with Ising\cite{Diep93},  XY and Heisenberg spins\cite{Delamotte2004,Loison}.
The cases of XY ($N=2$) and Heisenberg ($N=3$)
STA have been intensively studied since 1987\cite{kawamura87,kawamura88,azaria90,Loison94,Boub,Dobry,antonenko,Loison2000} , but only recently that the 20-year controversy comes to an end.\cite{tissier00b,tissier00,tissier01,itakura03,Peles,Kanki,Bekhechi,Zelli,Ngo08,Diep2008}
For details, see for example the review by Delamotte
et al\cite{Delamotte2004}.

The paper is organized as follows. Section II is devoted to the
description of the model and  the technical details of the Wang-Landau
(WL) methods as applied in the present paper.  Section III shows our
results.  Concluding remarks are given in section IV.

\section{Model and Wang-Landau algorithm}

\begin{figure}
\centerline{\epsfig{file=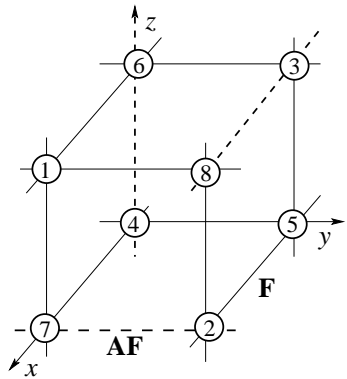,width=1.8in}} \caption{Fully frustrated simple cubic lattice.  Discontinued (continued) lines denote antiferromagnetic (ferromagnetic) bonds.} \label{fig:SCFF}
\end{figure}

We consider the FFSCL shown  in Fig. \ref{fig:SCFF}. The spins are
the classical XY model of magnitude
$S=1$. The Hamiltonian is given by
\begin{equation}\label{Ha}
{\cal H} = -\sum_{(i,j)} J_{ij}\mathbf{S}_i.\mathbf{S}_j,
\end{equation}
where $\mathbf{S}_i$ is the XY spin at the lattice site $i$, $\sum_{(i,j)}$ is made
over the NN spin pairs $\mathbf{S}_i$ and $\mathbf{S}_j$ with  interaction $J_{ij}$.  Hereafter we suppose that $J_{ij}=-J$ ($J>0$) for antiferromagnetic (AF) bonds indicated  by discontinued lines in Fig. \ref{fig:SCFF}, and $J_{ij}=J$ for ferromagnetic (F) bonds.  This model is a generalization in three dimensions (3D) of  the 2D Villain's model\cite{Villain} which has been extensively studied\cite{Berge,Lee91,Diep98}:  every face of the cube is frustrated because we know that a plaquette is frustrated when there is an odd number of AF bonds on its contour\cite{Diep2005,Villain}.  To describe the model, let us look first at the $xy$ plane (Fig. \ref{fig:SCFF}).  There, all interactions are F, except one AF line out of every two in the $y$ direction.  The same is for the $yz$ ($zx$) plane: one AF line of every two in the $z$ ($x$) direction.  Note that there is no intersection between AF lines.  Each plane is thus a 2D Villain's model.

\begin{figure}
\centerline{\epsfig{file=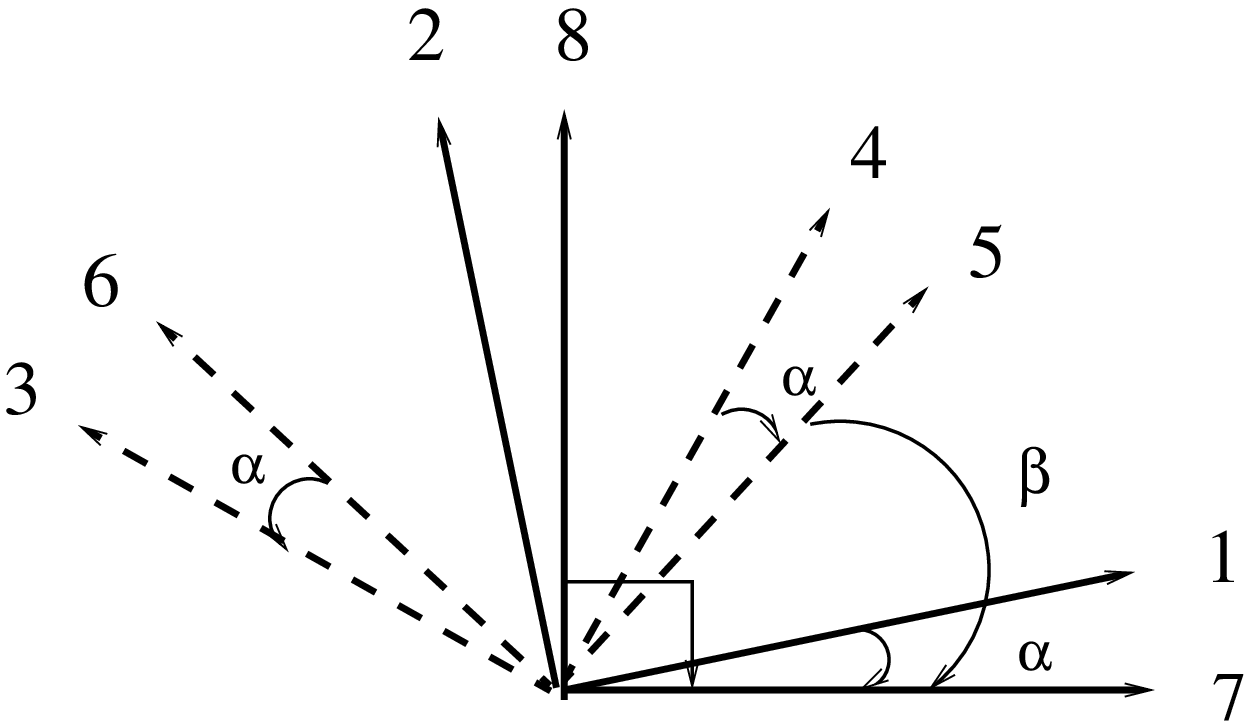,width=2.0in}}
\centerline{\epsfig{file=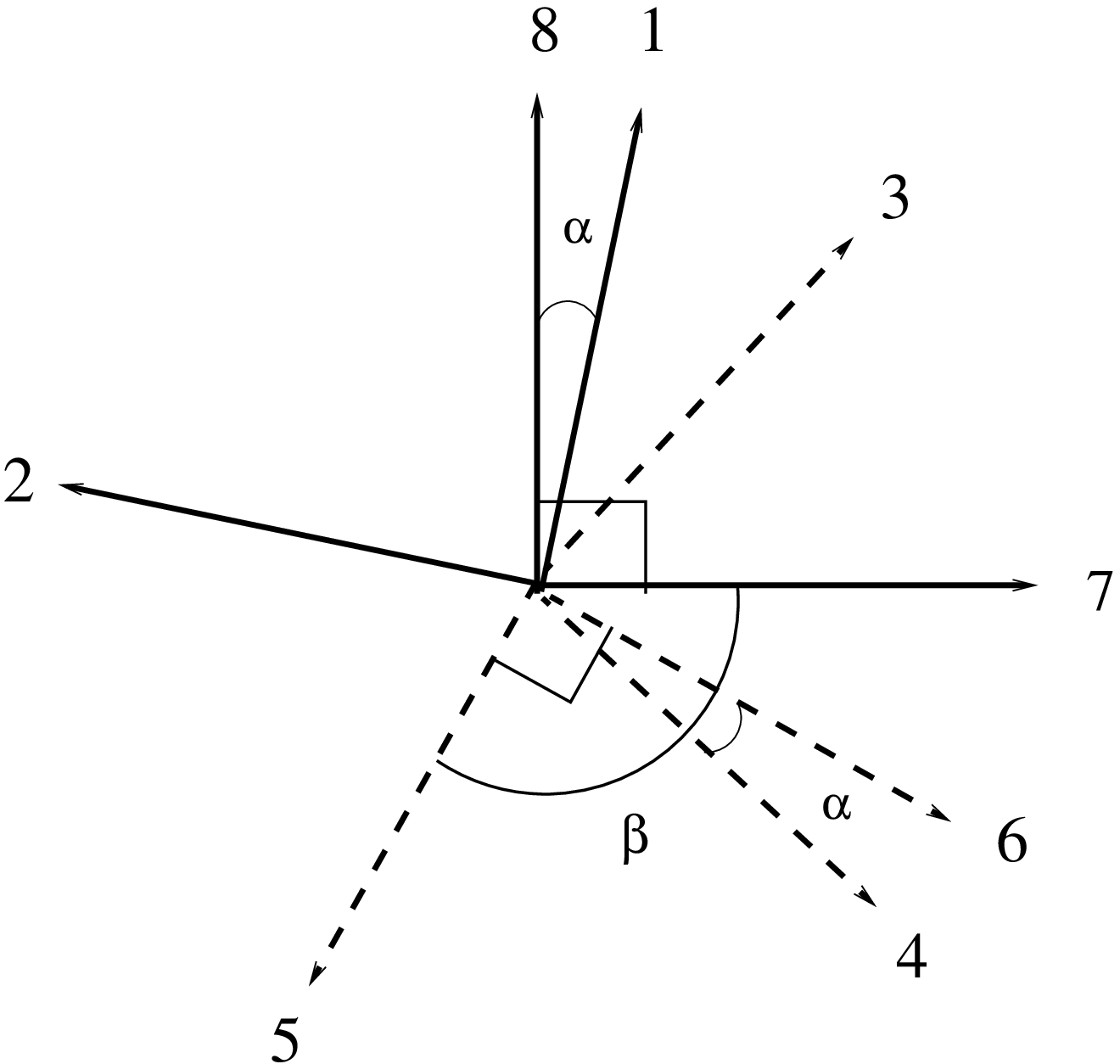,width=2.0in}}
\caption{Two of the 12 ground-state configurations of the fully frustrated simple cubic lattice. The numbers indicate the spins at the sites defined in Fig. 1. The angle $\alpha$ is $\alpha=\arccos(\frac{1+\sqrt{2}}{\sqrt {6}})$.  Upper: $\beta=\pi /4$, lower: $\beta=-3\pi /4$.} \label{fig:GS}
\end{figure}

Let us recall some results on the present model.  For the classical XY model on the FFSCL,  the GS is 12-fold degeneracy with non collinear spin configurations\cite{Diep85c}.  For convenience, let us define the local field acting on the spin $\mathbf {S}_i$ from its neighbors as
$\mathbf {h}_i=\sum_{j} J_{ij}\mathbf{S}_j$.   The GS can be calculated by noticing that the local field is the same at every site and is equal to $|\mathbf {h}|=2\sqrt {3}$ so that\cite{Diep85c}
\begin{eqnarray}
\mathbf {h}_5&=&2(\mathbf {S}_2+\mathbf {S}_3+\mathbf {S}_4)\\
\mathbf {h}_6&=&2(\mathbf {S}_1+\mathbf {S}_3-\mathbf {S}_4)\\
\mathbf {h}_7&=&2(\mathbf {S}_1-\mathbf {S}_2+\mathbf {S}_4)\\
\mathbf {h}_8&=&2(\mathbf {S}_1+\mathbf {S}_2-\mathbf {S}_3)
\end{eqnarray}
where the factor 2 results from the symmetric neighbors lying outside the cube and $J_{ij}=\pm 1$ depending on the bond has been used. Putting into square these equalities and using $\mathbf {h}^2=12$, $\mathbf {S}_i^2=1 (i=1,...,8)$, one has three independent relations which determine the relative orientation of every spin pair

\begin{eqnarray}
\mathbf {S}_2\cdot \mathbf {S}_3+\mathbf {S}_3\cdot \mathbf {S}_4+\mathbf {S}_2\cdot \mathbf {S}_4&=&0\\
-\mathbf {S}_1\cdot \mathbf {S}_3+\mathbf {S}_3\cdot \mathbf {S}_4+\mathbf {S}_1\cdot \mathbf {S}_4&=&0\\
\mathbf {S}_1\cdot \mathbf {S}_2+\mathbf {S}_2\cdot \mathbf {S}_4-\mathbf {S}_1\cdot \mathbf {S}_4&=&0
\end{eqnarray}
There are 12 solutions of these equations which can be described as follows\cite{Diep85c,Diep85b}. Consider just one of them shown in the upper figure of Fig. \ref{fig:GS}:  On a $yz$ face, the spins (displayed by continued vectors) on a diagonal are perpendicular, i.e. $\mathbf S_1 \bot \mathbf S_2$, $\mathbf S_7 \bot \mathbf S_8$. In addition, the
orthogonal dihedron  ($\mathbf S_1,\mathbf S_2$) makes an angle $\alpha=\arccos(\frac{1+\sqrt{2}}{\sqrt {6}})$ with the dihedron ($\mathbf S_7,\mathbf S_8$).
On the other $yz$ face the spins (displayed by discontinued vectors in Fig. \ref{fig:GS}) are arranged in the same manner: $\mathbf S_5 \bot \mathbf S_6$, $\mathbf S_4\bot \mathbf S_3 $  and the dihedron ($\mathbf S_4,\mathbf S_3$) makes an angle $\alpha$ with the dihedron ($\mathbf S_5,\mathbf S_6$).  Note that the dihedron ($\mathbf S_5,\mathbf S_6$) makes a turn angle $\beta=\pi /4$ with respect to the dihedron ($\mathbf S_7,\mathbf S_8$), and that the sum of the algebraic angles between spins on each face of the cube is zero.

There is another choice shown in the lower figure of Fig. \ref{fig:GS} where every  thing is the same as described above except $\beta=-3\pi/4$.  One has therefore two configurations with the choice of the $yz$ faces.  If one applies the same rule for the spins on the $xy$ faces and the $xz$ faces, one obtains in all 6 configurations.  Finally, together with their 6 mirror images, the total degeneracy is 12.\cite{Diep85b}

The above description of the GS shows an unusual degeneracy which can help to understand the first-order transition shown below by
relating this system to a Potts model where the GS degeneracy plays a determinant role in the nature of phase transition. Note however that all simulations have been carried out with the initial Hamiltonian (\ref{Ha}), no GS rigidity at finite temperature ($T$) has been imposed.

The first investigation of the nature of the phase transition of this model by the use of Metropolis Monte Carlo (MC) simulations has shown a second order transition with unusual critical exponents\cite{Diep85b}.  However, MC technique and computer capacity at that time did not allow to conclude the matter with certainty.
Recently, Wang and Landau\cite{WL1} proposed a MC algorithm for classical statistical models which allowed to study systems with difficultly accessed microscopic states. In particular, it permits to detect  with efficiency weak first-order transitions\cite{Ngo08,Diep2008} The algorithm uses a random walk in energy space in order to obtain an accurate estimate for the density of states $g(E)$ which is defined as the number of spin configurations for any given $E$. This method is based on the fact that a flat energy histogram $H(E)$ is produced if the probability for the transition to a state of energy $E$ is proportional to $g(E)^{-1}$.

We summarize how this algorithm is implied here. At the beginning of the simulation, the density of states (DOS) is set equal to one for all possible energies, $g(E)=1$.
We begin a random walk in energy space $(E)$ by choosing a site randomly and flipping its spin with a probability
proportional to the inverse of the temporary density of states. In general, if $E$ and $E'$ are the energies before and after a spin is flipped, the transition probability from $E$ to $E'$ is
\begin{equation}
p(E\rightarrow E')=\min\left[g(E)/g(E'),1\right].
\label{eq:wlprob}
\end{equation}
Each time an energy level $E$ is visited, the DOS is modified by a modification factor $f>0$ whether the spin flipped or not, i.e. $g(E)\rightarrow g(E)f$.
  At the beginning of the random walk, the modification factor $f$ can be as large as $e^1\simeq 2.7182818$. A histogram $H(E)$ records the number of times a state of energy $E$ is visited. Each time the energy histogram satisfies a certain "flatness" criterion, $f$ is reduced according to $f\rightarrow \sqrt{f}$ and $H(E)$ is reset to zero for all energies. The reduction process of the modification factor $f$ is repeated several times until a final value $f_{\mathrm{final}}$ which close enough to one. The histogram is considered as flat if
\begin{equation}
H(E)\ge x\%.\langle H(E)\rangle
\label{eq:wlflat}
\end{equation}
for all energies, where $x\%$ is chosen between $70\%$ and $95\%$
and $\langle H(E)\rangle$ is the average histogram.

The thermodynamic quantities\cite{WL1,brown} can be evaluated by
%\begin{eqnarray}
$\langle E^n\rangle =\frac{1}{Z}\sum_E E^n g(E)\exp(-E/k_BT)$,
$C_v=\frac{\langle E^2\rangle-\langle E\rangle^2}{k_BT^2}$,
$\langle M^n\rangle =\frac{1}{Z}\sum_E M^n g(E)\exp(-E/k_BT)$, and
$\chi=\frac{\langle M^2\rangle-\langle M\rangle^2}{k_BT}$,
%\end{eqnarray}
where $Z$ is the partition function defined by
%\begin{equation}
$Z =\sum_E g(E)\exp(-E/k_BT)$.
%\label{eq:partfunc}
%\end{equation}
The canonical distribution at any temperature can be calculated simply by
%\begin{equation}
$P(E,T) =\frac{1}{Z}g(E)\exp(-E/k_BT)$.
%\label{eq:pe}
%\end{equation}

In this work, we consider a energy range of interest\cite{Schulz,Malakis}
$(E_{\min},E_{\max})$. We divide this energy range to $R$ subintervals, the minimum energy of each subinterval is $E^i_{\min}$ for $i=1,2,...,R$, and maximum of the subinterval $i$ is $E^i_{\max}=E^{i+1}_{\min}+2\Delta E$,
where $\Delta E$ can be chosen large enough for a smooth boundary between two subintervals. The WL
algorithm is used to calculate the relative DOS of each subinterval $(E^i_{\min},E^i_{\max})$ with the
modification factor $f_\mathrm{final}=\exp(10^{-9})$ and flatness criterion $x\%=95\%$.
We reject the suggested spin flip and do not update $g(E)$ and the energy histogram $H(E)$ of
the current energy level $E$ if the spin-flip trial would result in an energy outside the energy segment.
The DOS of the whole range is obtained by joining the DOS of each
subinterval $(E^i_{\min}+\Delta E,E^i_{\max}-\Delta E)$.

\section{Results}

 We used the system size of $N\times N \times N$ where
$N$ varies from 24 up to 48. We stop at $N=48$ because, as seen below, the transition at this size shows a definite answer to the problem studied here.  Periodic
boundary conditions are used in the three directions.  $J=1$ is
taken as the unit of energy in the following.

\begin{figure}
\centerline{\epsfig{file=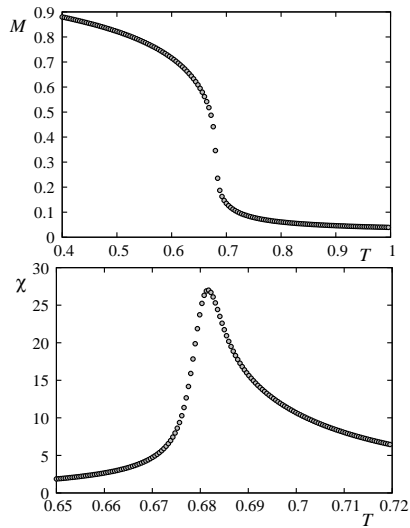,width=2.1in}} \caption{Magnetization and susceptibility
 versus $T$ for $N=$24.} \label{fig:MX}
\end{figure}

\begin{figure}
\centerline{\epsfig{file=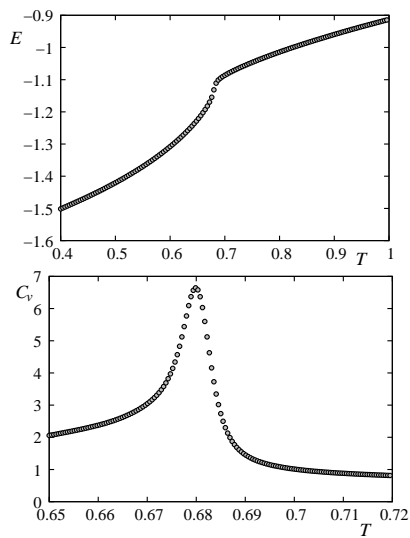,width=2.1in}} \caption{Energy per spin and specific heat
 versus $T$ for $N=$24.} \label{fig:EC}
\end{figure}

\begin{figure}
\centerline{\epsfig{file=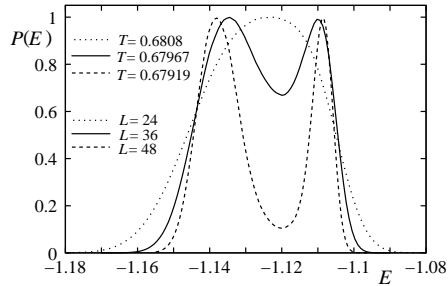,width=2.3in}} \caption{Energy
histogram for several sizes $N=24$, 36, 48 at $T_c$ indicated on the
figure.}\label{fig:PE}
\end{figure}

We show in Fig. \ref{fig:MX} the magnetization and the susceptibility
and in Fig. \ref{fig:EC} the energy per spin and the specific heat, for $N=24$.  All these quantities show a transition with an second-order aspect.   However, we know that many systems show a first-order transition only at very large sizes.  This is indeed the case.
The energy histograms obtained by WL technique for three representative sizes $N=24$, 36 and 48 are shown in Fig. \ref{fig:PE}. As seen, for $N=24$, the energy histogram, though unusually broad, shows a single peak indicating a continuous energy at the transition as observed before in Fig. \ref{fig:EC}. The double-peak histogram starts already at $N=36$ and the dip between the two maxima becomes deeper with increasing size, as observed at $N=48$.  We note
that the distance between the two peaks, i. e.
the latent heat, increases with increasing size and reaches $\simeq 0.03$ for $N=48$.  This is rather large compared
with the value $\simeq 0.009$ for $N=120$ in the XY STA\cite{itakura03,Peles,Kanki,Ngo08} and with 0.0025 for $N=150$ in the Heisenberg case\cite{Diep2008}.
We give here the values of $T_c$ for a few sizes:  $T_c =
0.68080\pm0.00010$, $0.67967\pm 0.00010$ and $0.67919\pm 0.00010$ for $N=24$, 36 and 48,
respectively.

Note that the double-peak structure is a sufficient
condition, not a necessary condition in old-fashion MC simulations (i. e. not WL method), for a first-order transition.   This is because in old-fashion MC simulations performed at a given $T$, we often encounter the situation where, at the transition, the system is spatially composed of two (or more) parts:  the ordered phase with energy $E_1$ and the disordered phase with energy $E_2$.  Since in  old-fashion MC simulations, we make histogram from the total system energy, i. e.  $E_1+E_2$, the histogram will record the 'average' energy $E_1+E_2$, therefore the double peak structure will not be seen.  Such a coexistence at any time of the ordered and disordered phases in some first-order transitions makes it impossible in old-fashion MC simulations to detect two peaks.  In our present WL flat-histogram method, the double-peak structure is obtained from the DOS histogram which gets rid of the problem of spatial coexistence of the two phases discussed above.  Therefore, the double-peak structure is a necessary condition for a first-order transition as it should be.

Let us say a few words on the correlation length. It is known that the correlation length is finite at the transition point in a first-order transition.  For very strong first-order transitions, this correlation length is short so that the first-order character is detected in simulations already at small lattice sizes.  For weak first-order transitions, the correlation length is very long. To detect it one should study very large lattice sizes as in the present paper.   Direct calculation of the correlation length is not numerically easy.  Fortunately, one has other means such as the WL method to detect more easily weak first-order transitions.

%\begin{figure}
%\centerline{\epsfig{file=STA36PE.eps,width=2.8in}} \caption{Energy
%histogram for $N=36$.} \label{fig:STA36PE}
%\end{figure}

%\begin{figure}
%\centerline{\epsfig{file=STA36GE.eps,width=2.8in}}
%\caption{Density of state for $N=36$.} \label{fig:STA36GE}
%\end{figure}

%\begin{figure}
%\centerline{\epsfig{file=STAH3CV.eps,width=2.5in}} \caption{Specific heat
% versus $T$  for  $N=96, 120, 150$.}
%\label{fig:STAH3CV}
%\end{figure}

To close this section, let us emphasize two points: i) First, the first-order transition observed here may come from the fact that the GS of the present XY FFSCL model has a 12-fold degeneracy which is reminiscent of the 12-state Potts model.  In three dimensions, the latter model has a first order transition. Note however that this conclusion is not always obvious because the continuous degrees of the order parameter mask in many cases the symmetry argument based on discrete models\cite{Diep98}, ii) Second, some other XY frustrated systems such as the FCC\cite{Diep89}, HCP\cite{Diep92} and helimagnetic\cite{Diep89a} antiferromagnets show also a first-order transition in MC simulations.  Though the nonperturbative renormalization group has been extensively used for the STA case\cite{Delamotte2004} due to its long-lasting controversy, we believe that the other cases are worth to study in order to verify that the validity of that theory is not limited to the STA but is universal for frustrated systems of vector spins.

\section{Concluding Remarks}

Using the powerful WL flat histogram technique, we have studied the phase transition in the XY fully frustrated simple cubic lattice. The
technique is very efficient because it helps to overcome extremely long
transition time between energy valleys in systems with a
first-order phase transition.   We found that the transition is clearly of first-order at large lattice sizes in contradiction of early studies using standard MC algorithm and much smaller sizes\cite{Diep85b}. The result presented here will serve as a testing ground for theoretical methods such as the renormalization group which still has much difficulty to deal with frustrated systems\cite{Delamotte2004}.

One of us (VTN) would like to thank  the University of Cergy-Pontoise for a financial support during the course of this work. He is grateful to Nafosted of Vietnam National Foundation
for Science and Technology Development, for support (Grant No. 103.02.57.09). He also thanks the NIMS (National Institute for Mathematical
Sciences, Korea) for hospitality and financial support.

{}

\end{document}